# Use of Groundwater Lifetime Expectancy for the Performance Assessment of a Deep Geologic Radioactive Waste Repository: Application to a Canadian Shield Environment

Y.-J. Park[1*], F. J. Cornaton[1,3], S. D. Normani[2],

J. F. Sykes[2], and E. A. Sudicky[1]

1: Department of Earth and Environmental Sciences
University of Waterloo
Waterloo, Ontario, Canada N2L 3G1
E-mail: yj2park@sciborg.uwaterloo.ca,
sudicky@sciborg.uwaterloo.ca,
Fax: 1-519-746-7484

2: Department of Civil and Environmental Engineering
University of Waterloo
Waterloo, Ontario, Canada N2L 3G1
E-mail: sdnorman@uwaterloo.ca,
sykesj@uwaterloo.ca

3: Now at:
CHYN, Centre of Hydrogeology, University of Neuchâtel
Emile-Argand 11, CP 158
CH-2009 Neuchâtel, Switzerland
fabien.cornaton@unine.ch

* Corresponding author

# Abstract

*Cornaton et al.* [2007] introduced the concept of lifetime expectancy as a performance measure of the safety of subsurface repositories, based upon the travel time for contaminants released at a certain point in the subsurface to reach the biosphere or compliance area. The methodologies are applied to a hypothetical but realistic Canadian Shield crystalline rock environment, which is considered to be one of the most geologically stable areas on Earth. In an approximately $10 \times 10 \times 1.5$ km$^3$ hypothetical study area, up to 1000 major and intermediate fracture zones are generated from surface lineament analyses and subsurface surveys. In the study area, mean and probability density of lifetime expectancy are analyzed with realistic geologic and hydrologic shield settings in order to demonstrate the applicability of the theory and the numerical model for optimally locating a deep subsurface repository for the safe storage of spent nuclear fuel. The results demonstrate that, in general, groundwater lifetime expectancy increases with depth and it is greatest inside major matrix blocks. Various sources and aspects of uncertainty are considered, specifically geometric and hydraulic parameters of permeable fracture zones. Sensitivity analyses indicate that the existence and location of permeable fracture zones and the relationship between fracture zone permeability and depth from ground surface are the most significant factors for lifetime expectancy distribution in such a crystalline rock environment. As a consequence, it is successfully demonstrated that the concept of lifetime expectancy can be applied to siting and performance assessment studies for deep geologic repositories in crystalline fractured rock settings.

# 1. Introduction

Crystalline rock is considered to have among the lowest permeability and porosity of all geologic materials (as it is often considered to be impermeable bedrock for shallow groundwater flow) and it has led to the suggestion that repositories for highly toxic wastes such as spent nuclear fuel could be safely hosted at depth in such environments [*Witherspoon et al.*, 1981]. However, it is also well known that flow systems in fractured porous media can be highly influenced by interconnected permeable fracture networks, acting as pathways for contaminant migration, thereby deteriorating the long term performance and safety of subsurface repositories [*Bear et al.*, 1993; *National Research Council*, 1996; *Adler and Thovert*, 1999].

Safety or risk for subsurface repositories is assessed using criteria such as the travel time of released contaminants from the waste site to compliance boundaries, the dose or the degree of dilution of the contaminants when reaching the biosphere, and the uncertainty for the prediction. In this context, *Cornaton et al.* [2007] introduced the concept of lifetime expectancy, which is defined as the time that water molecules spend in the subsurface after release from a waste site and before they reach an exit boundary. It was also demonstrated that mean and probability density for lifetime expectancy could be used to define potentially optimal locations for underground repositories in fractured porous media in terms of the longest travel time from the waste site to the biosphere.

A subsurface waste repository facility should be located in a region that is geologically stable and likely to remain stable, especially for long-term performance [*AECB*, 1987]. Geological stability is a characteristic that would clearly enhance the protective function of a geologic barrier and it could also enhance the long-term predictability of conditions in the repository. The plutonic rock of the Canadian Shield, one of the most geologically stable areas on Earth, could be considered to provide the requisite stable environment for a geologic barrier. *Sykes et al.* [2003] analyzed the regional scale flow system in a hypothetical plutonic Canadian Shield setting (approximately 6000 km$^2$ in area and about 1.5 km deep), where they found that an essential requirement of the analysis of regional-scale groundwater flow in crystalline rock is the preservation and accurate description of the complex topography, hydraulic properties distribution such as the location of permeable fracture zones and the relationship between fracture zone permeability and depth, the long-term influence of shield brine and the assessment of glacial impact.

In this study, among others, we focus on the effects of the geometric and hydraulic characteristics of the fracture zones on the lifetime expectancy distribution in a sub-regional scale domain (a smaller sub-catchment in the regional domain analyzed by *Sykes et al.* [2004]), where up to 1000 major and intermediate fracture zones could exist, as revealed by surface lineament analyses and subsurface surveys [*Srivastava*, 2002]. Following the methodologies suggested by *Cornaton et al.* [2007] and in order to analyze the suitability for hosting deep geologic repositories, we apply the theory and the numerical model to a crystalline fractured rock environment in a hypothetical but realistic plutonic crystalline Canadian Shield setting. We also analyze the uncertainties from various sources such as the location of fracture zones, fracture zone permeability variation with depth, the permeability-porosity relationship, and fracture zone width distribution, to ensure the applicability and the utility of the theory and the numerical model.

## 2. Crystalline Environment of the Canadian Shield

### 2.1. Geological Background

The Canadian Shield is the largest area of Archean rocks in the world, more than 2.5 billion years old. Most of the Canadian Precambrian Shield consists of granitic rocks and gneiss laced with sinuous greenstone volcanic belts and broader areas of sedimentary rocks. The Canadian Shield is subdivided into major geologic units called provinces or orogens on the basis of the structure, history of deformation, and the estimated age of formation of the rock. The Superior Province of the Canadian Shield is the largest exposure of Archean rock on Earth and it consists of east-west trending belts of metavolcanic, metasedimentary, and plutonic rocks. It comprises most of the Shield in Ontario and was formed between 3.1 and 2.7 billion years ago. The last deformation occurred around 2.5 billion years ago and the rock was repeatedly faulted and locally intruded during the Proterozoic period (2.5 billion to 570 million years ago) [*Encyclopedia Britannica*, 2006; *Davison et al.*, 1994].

The land surface of the Canadian Shield is currently rising slowly as a result of the melting and retreat of the Laurentide Ice Sheet about 10,000 years ago [*Tarasov and Peltier*, 1999; *Peltier*, 2002]. During a glaciation, the surface is depressed due to the mass of the overlying ice. The rate of rebound decreases with time, but it is still occurring at a rate of about 6 m per 1000 years near Hudson Bay in northeastern Canada. The rate of rebound declines with distance away from Hudson Bay and approaches zero at the southern edge of the Shield [*Davison et al.*, 1994].

## 2.2. Shield Hydrogeology

In general, shallow groundwater flow patterns are generated by the topography of the watershed and modified by the spatial variations in the permeability of the subsurface lithologic units, by the presence of faults and fracture zones, and by density differences in the resident groundwater. For watersheds in the Canadian Shield, recharge and downward groundwater movement occurs at topographically high regions, upward groundwater flow and discharge occurs at topographic depressions such as stream valleys or lake basins. Groundwater through-flow that is sub-parallel to the land surface occurs under medium elevations between mounds and vales [*Toth and Sheng*, 1996].

The correlation between topography and groundwater flow patterns has been investigated in several crystalline rock settings. The analysis of water levels in boreholes at the Swedish Äspö Hard Rock Laboratory indicates that the undisturbed water table approximately follows the topography and that annual fluctuation in the water table is greater for areas near surface water divides than for areas close to discharge areas [*Stanfors et al.*, 1999]. It was also observed that the location of the water table is locally irregular because the hydraulically active fractures are sparsely distributed and not well connected hydraulically. At depth, the irregular pressure distribution becomes smoother and to a large extent, becomes governed by the fracture zones with higher permeabilities. Similar observations were made at the Atomic Energy of Canada Limited (AECL)'s Whiteshell Research Area (WRA), Manitoba [*Stanchell et al.*, 1996]

The salinity of groundwater generally increases with depth in the plutonic rock on the Canadian Shield. Two theories have been postulated to explain the presence of the highly saline (up to 1.3 kg/L of density) groundwater in the deeper rock. Salinity may have originated from groundwater recharge during episodes of marine intrusion, and alternatively, it may be a result of rock-water interactions [*Clark et al.*, 2000; *Bottomley et al.*, 2003; *Gascoyne*, 2004]. Shallow recharging groundwater typically has a low content of dissolved solids and is of a calcium bicarbonate composition. With the slow movement of the infiltrating water through fractures, surfaces of open fractures are altered to clay minerals and calcium carbonate is precipitated [*Gascoyne*, 2004].

## 3. Sub-Regional Domain and Numerical Model

### 3.1. Sub-Regional Hypothetical Domain

*Sykes et al.* [2003] analyzed the regional groundwater system in a typical shield environment, based on the data collected at the Whiteshell Research Area (WRA) and the Underground Research Laboratory (URL) in Manitoba. Within the regional domain studied by *Sykes et al.* [2003], the water table was found to be a subdued replica of the topographic surface, and the water table divides roughly corresponded with the surface water catchment boundaries. A sub-regional domain – a sub-catchment within the regional domain studied by *Sykes et al.* [2003] – was selected to represent the details of the typical plutonic Canadian Shield environment (Figure 1) [*Sykes et al.,* 2004]. Note that the analyses in the sub-regional domain can be considered to be realistic in the sense that the topography, lineament analysis, and fracture network generation are performed in the existing site but it is also hypothetical because typical hydrogeological settings in the Canadian Shield environment, observed in WRA and URL, are assumed for the distribution of hydraulic properties in fracture zones and matrix blocks. The domain has an areal dimension of approximately 100 km$^2$, an easting extent of about 11 km and a northing extent of about 12 km. The sub-regional domain has low topographic relief, with surface elevations ranging from 350 m to 410 m, and includes wetlands and lakes along with creeks, streams, and rivers, many of which are linear and suggestive of the surface expression of major fracture zones [*Srivastava*, 2002].

*Srivastava* [2002] developed a discrete-fracture network model, which provides geostatistical tools for the probabilistic simulation of the propagation of surface lineaments to depth in order to build geologically and geomechanically plausible fracture networks that honor the type of information typically available from non-invasive site characterization investigations. The model utilizes such information as the surface expression of some of the fractures, which manifest themselves as lineaments on aerial photographs and remote sensing images, statistics on fracture density and length distributions, general structural geology principles that describe the down-dip behavior of fractures, and truncation rules that arise from regional tectonic and geologic considerations. Additional surface lineaments were created to account for the extension of existing major lineaments, and to increase the fracture density in areas where overburden cover would have obscured the surface lineaments or where aerial photography had weak contrast (Figure 2a).

## 3.2. Numerical Simulator and Model Development

FRAC3DVS is a variably-saturated groundwater flow and reactive solute transport simulator designed for application to discretely fractured porous media [*Therrien and Sudicky*, 1996; *Therrien et al.*, 2003]. It can also simulate flow and transport in zoned-type porous or dual porosity/dual permeability media in which the material property contrasts between various zones is large because of advanced numerical solution strategies employed (see *Therrien et al.* [2003] for details). Backward-in-time travel time analysis framework was implemented recently in the simulator and applied in this study. A discrete fracture dual permeability conceptualization is adapted for the simulations carried out for sub-regional domain, where fracture zones as two-dimensional plane features represent preferential pathways for flow and transport and the three-dimensional matrix domain is for less permeable fractured rock.

The three-dimensional sub-regional domain was discretized into approximately 850,000 nodes and 790,000 brick elements, with each element having a planimetric dimension of 50 m by 50 m and 19 layers of 10 m (near surface) to 200 m (at depth) thickness. In the model, the first 5 layers from surface have variable thickness to adjust the elevation of ground surface and the remaining layers have constant thickness.

The geometry of curve-planar fracture zone networks, developed by *Srivastava* [2002] (Figure 2a), was mapped onto orthogonal faces of brick elements (Figure 2b) as permeable two-dimensional planes [*Therrien et al.*, 2003]. Figure 2 shows that the fracture planes mapped onto elemental faces could accommodate both the dip and orientation of original curve-planar fracture zones in the scale under consideration, despite their stepped nature [*Sykes et al.*, 2004].

The porosity of the rock matrix was assumed to be uniformly distributed throughout the modeling domain and to have a value of 0.003 [*Sykes et al.*, 2003, 2004]. The horizontal matrix permeability is set to be $10^{-13}$ m$^2$ near the surface, exponentially decreasing with depth to $10^{-19}$ m$^2$ at about 700 m below ground surface, and remains constant for deeper layers. The higher permeability values for the shallow depths represent the presence of moderately fractured rock while the lower values at greater depths reflect the dominance of sparsely-fractured rock. Vertical matrix permeability is assumed to be greater than the horizontal permeability by an order of magnitude down to about 300 m below ground surface, reflecting the dominance of vertical or sub-vertical fractures in the region [*Everitt*, 2002].

A statistical model was developed to describe the relationship between depth and fracture zone permeability, based on the data collected from WRA (Figure 3) [*Normani et al.*, 2007]. The essence of the data provided in Figure 3a is that the permeability has a decreasing tendency with depth in general but it varies widely; thus, fracture zones could remain permeable at depth with a certain finite probability. Moving-average percentile analysis can account for both general tendency and probabilistic characteristics for the relationship.

It is shown in Figure 3 that most of measured fracture-zone permeability values range from $10^{-12}$ to $10^{-15}$ m$^2$ at ground surface, decrease by two to three orders of magnitude down to about 600 m below ground surface, and become statistically homogeneous, ranging from $10^{-13}$ to $10^{-18}$ m$^2$ at depth. On average (50th percentile in Figure 3b), fracture-zone permeability is about $10^{-13}$ m$^2$ at ground surface, mildly decreases to 200 m below ground surface, rapidly decreases by about three orders to 500 m below ground surface, and becomes $10^{-16}$ m$^2$ at depth.

Hydraulic properties of fractured porous media such as permeability and porosity can be derived, based upon the parallel plate flow concept and observations of fracture patterns [*Snow*, 1968, 1969; *Bear*, 1972; *Chen et al.*, 1999]. For a homogeneous isotropic fractured system, fracture zone porosity ($n_{bulk}$) can be represented as a function of fracture zone permeability ($k_{bulk}$), fracture spacing or density ($\rho$), and matrix porosity ($n_m$).

$$n_{bulk} \approx 3\left(6 k_{bulk} \rho^2\right)^{1/3} + n_m \qquad (1)$$

When a fracture density of 10 m$^{-1}$ is used to compute the fracture zone porosity for given fracture zone permeability values provided in Figure 3b, $n_{bulk}$ ranges from 0.003 to 0.01. This agrees reasonably well with measured values.

A log-normal distribution for fracture zone width was derived from the data collected at the WRA:

$$F_X(x) = \frac{1}{x\sigma\sqrt{2\pi}} e^{\left(-\frac{(\ln(x/m))^2}{2\sigma^2}\right)}, \quad \text{where } \sigma = 0.48, m = 3.27 \qquad (2)$$

Observed average fracture-zone width was about 3.3 m and most were narrower than 10 m. In the model, each fracture zone is assumed to have a constant width but the value could be allowed to vary from one zone to another with probability, as given in Equation (2). Note that the fracture zone propagation model includes termination rules and the decreasing trend in width with depth is implicitly incorporated by the model.

## 4. Groundwater Flow and Lifetime Expectancy

### 4.1. Groundwater Flow

In order to understand the general dynamics of groundwater flow in a Shield environment, only saturated groundwater flow is assumed in the computational domain. This is justified by the generally shallow depths to the water table in the Canadian Shield setting. As a reference case, hydraulic properties in fracture zones are assumed such that the permeability follows the average depth-permeability relationship shown in Figure 3b (50th percentile or median trend), the porosity is computed as a function of permeability with a fracture density of 10 $m^{-1}$ and the matrix porosity of 0.003, and the width follows a log-normal distribution from the Equation (2). A specified head boundary condition was applied to represent observed surface water bodies, including lakes, wetlands, and rivers (Figure 1) and a recharge rate of 1.0 mm/year was applied to the remaining top surface of the model. The recharge rate was estimated by applying a specified head boundary condition equivalent to surface elevation to the entire top surface of the model and averaging total recharge into the domain over the top surface. The distribution of recharge and discharge areas over the ground surface can be irregular, depending on the resolution of the digital elevation model used for the domain; when the specified head boundary condition was applied to the entire top surface, irregularities in the computed surface discharge were apparent, while the combined first- and second-type boundary conditions eliminated these irregularities. The bottom of the model was impermeable assuming no-flow as are side boundaries.

The simulated head distribution in the domain indicates that groundwater generally flows from the topographically higher northern region towards the south and that head distributions can be more irregular near surface than at depth due to irregular distributions in ground surface elevations, surface water bodies, and major fracture zones (Figure 4). The results agree well with the theory and the observations that the vertical location of the water table is closely correlated to ground surface elevation and pressure is observed to become smoother at depth, although the water table location could be irregular due to the presence of permeable fracture zones [*Toth and Sheng*, 1996; *Stanfors et al.*, 1999; *Stanchell et al.*, 1996].

The calculated fluid flux distribution provided in Figure 5 confirms that fracture zones can act as permeable pathways for fluid flow and contaminant migration even at depths of 1 km below ground surface. It is clear from Figure 5 that groundwater flow is more active near surface than at depth because the permeabilities in the fracture zones and matrix blocks are higher and the hydraulic driving force is stronger near the surface. Note, however, that major fracture zones may remain hydraulically active at depth, although fracture zones become sparser and less permeable at depth and groundwater flow becomes relatively stagnant.

**4.2. Lifetime Expectancy**

Mean lifetime expectancy (MLE) – the average time a water molecule will spend after release from a certain point and before reaching exit boundaries – was computed throughout the domain following the theory suggested by *Cornaton et al.* [2007] (Figure 6). It is worth noting that the MLE is described by a steady-state backward-in-time transport equation (Equation (9) in *Cornaton et al.* [2007]) and it can be solved throughout the domain at once although forward-in-time transport equation can also be solved to simulate the transport of MLE from each source location to biosphere as many times as the number of potential sources. As the simulations are aimed at searching an optimal location for repositories, backward-in-time framework is not only computationally more efficient but also describes the nature of the problem better. It is shown in Figure 6 that MLE values range over more than 5 orders of magnitude; it generally increases as the release point becomes deeper, it is smallest in major fracture zones, and it increases from the locations of fracture zones towards the middle of the matrix blocks. Although there is a tendency that MLE values are less in the region below the northern recharge area than the region below the southern discharge area, this tendency may not be a dominant pattern for the distribution of MLE throughout the domain. It is implied from the results shown in Figure 6 that the safety for subsurface repositories can strongly depend on depth, and the existence and location of fracture zones, but the relative location in the regional groundwater flow system (recharge or discharge areas) can be less significant as indicated by *Cornaton et al.* [2007].

Figure 7 shows cumulative relative frequency plots for MLE at four elevations within the domain in Figure 6. The frequency plots were generated by spatially integrating the computed MLE distributions at four elevations. The results indicate that MLE is strongly dependent on depth, with greater variability at shallower depths due to the increased density of the more permeable fracture zones, but its variability decreases in the deeper diffusion-dominated transport environment.

MLE is the first temporal moment of lifetime expectancy probability and, thus, a greater MLE does not always guarantee a lesser dose, or later arrival of contaminants at the biosphere, because the averaged travel time of contaminants can be much longer than first or peak arrivals in multiple-pathway flow system, even though it might be perceived to be the case [*Cornaton et al.*, 2007]. *Cornaton et al.* [2007] clearly demonstrated, with illustrative fractured porous-media examples, that it is more desirable to compute the probability density function for lifetime expectancy, in order to better analyze the suitability of a geologic setting to host a subsurface repository. Although it is possible to compute the mass fluxes expected at the biosphere, as a result of any given release function at any given volume or any union of subsets of volume (see Equations 6 and 10 in *Cornaton et al.*, 2007), four specific release locations were selected from Figure 6 (P1 – P4) for illustration and comparison. Each location has an approximate 100m by 200m by 1km dimension and the lifetime expectancy probability was simulated for 2 million years. For given release locations P1 and P2, the mean for lifetime expectancy was calculated to be approximately $10^5$ and $10^6$ years, respectively, but the first peak may arrive at the discharge area at approximately $2\times10^4$ and $10^5$ years (Figure 8). For the deeper release locations (P3 and P4), MLE was greater than $10^6$ and $10^7$ years and first peak does not arrive at the surface discharge area for $2\times10^6$ years. This result implies that strong tailing for transport in fractured porous media due to multiple pathways and matrix diffusion can result in significant differences between the first temporal moment and the first peak arrival time, but longer MLE would still be an indicator for later arrival of the water molecules, as indicated by *Cornaton et al.* [2007].

## 5. Uncertainty Analysis

The possible presence of relatively permeable fracture zones have attracted concern for the safety of subsurface repositories because they could serve as the principle pathways for contaminant migration. It is also shown in this study and in *Cornaton et al.* [2007] that lifetime expectancy as a safety indicator is strongly influenced by the presence of such fracture zones. However, it is unlikely that the locations and hydraulic properties of fracture zones can be characterized deterministically, but they, at best, can be characterized statistically with a certain degree of uncertainty (see Section 3.2. Numerical Model Development). In this section, we analyze the uncertainty in lifetime expectancy associated with the uncertainties in fracture zone location and fracture zone properties.

### 5.1. Hydraulic Properties for Fracture Zones

In order to investigate the effects of fracture-zone permeability distributions on groundwater flow in a Shield environment, mean lifetime expectancy is simulated for different permeability-depth relationships, such as uniform permeability and a probabilistic decrease of permeability with depth, and is compared to the MLE distribution for a deterministic decrease of permeability with depth. Results are in Figures 6 and 7. For the uniform permeability model, an average permeability value measured near ground surface ($10^{-13}$ m$^2$ in Figure 3) is assigned for all the fracture zones in the domain. For the probabilistic model, the depth-permeability relationship in each fracture zone follows an $i^{th}$ percentile in Figure 3b, where $i$ is a random number uniformly distributed between 1 and 100, and thus the relationships are different among fracture zones.

In Figure 9, MLE distributions are compared for a uniform fracture zone permeability of $10^{-13}$ m$^2$, and for a probabilisitically decreasing permeability with depth, according to the statistical model illustrated in Figure 3b. Compared to Figure 6, MLE within the fracture zones is notably decreased for the uniform permeability case because the fracture-zone permeability becomes smaller at depth. The deterministic and probabilistic depth-permeability models share a similar pattern in MLE distribution, even though the distribution is less uniform at each elevation. Cumulative frequency plots are also compared to the reference average deterministic model in Figure 10. The results in Figure 10a clearly show that with the uniform model, leading tails for smaller MLE become stronger at deeper elevations, due to a higher contrast in permeability between the fracture zones and the matrix. With the probabilistic model, fracture-zone permeability can vary among fracture zones, leading to wider MLE distributions at each elevation (Figure 10b), but its influence is minor for the given statistical model. The results shown in Figures 9 and 10 indicate that a depth-permeability relationship can play a critical role as a higher fracture-zone permeability could diminish the MLE, but the influence of a probabilistic relationship is relatively limited for the lifetime expectancy in the matrix blocks.

The relationship between the permeability and the porosity of a fracture zone in

Equation (1) provides estimates for fracture-zone porosity that can range from 0.32 to 0.56 % for given permeability values for the reference case, with a fracture density of $10^{-1}$ m. Although a narrow range of porosity values is used for the reference simulation, it could be significantly higher in some fracture zones, depending on their origin. In order to evaluate the influence of fracture-zone porosity on MLE distribution in a Shield environment, uniform constant porosity values of 0.3 and 3% are considered as lower and upper limits for the simulations, while keeping other parameters unchanged from the reference simulation. Figure 11 compares cumulative frequency plots of MLE at four elevations within the domain when the uniform porosity models are utilized. The results provided in Figure 11 indicate that the effect of fracture-zone porosity on MLE is negligible.

*Cornaton et al.* [2007] suggested analytical solutions for MLE in a fractured porous system as follows:

$$E_f(z) = \frac{z}{2bq_f}\left(\phi_m L + 2b\phi_f\right) \quad (3a)$$

$$E_m(x,z) = -\frac{1}{2D_m}x(x-L) + E_f(z) \quad (3b)$$

where $x$ ($[0,\infty)$) and $z$ ($[0,L]$) are spatial coordinates along and perpendicular to the equally-spaced parallel vertical fractures of aperture $2b$, and where $q_f$ is the upward fluid flux along each fracture, $E$ is MLE, $\phi$ is porosity, $D$ is effective dispersion coefficient, and subscripts $f$ and $m$ denote fracture and matrix respectively. It is implied in (3) that MLE in the fracture and matrix domain may not be influenced by fracture zone porosity

($\phi_f$) if the matrix pore volume is much larger than the fracture pore volume ($\phi_m L \gg 2b\phi_f$), as is often the case in crystalline rocks settings.

A change in the width and transmissivity of a fracture zone results in a change in fluid flux along the fracture zone ($q_f$ in (3)). Fracture zone widths measured at the WRA indicated width follows a log-normal distribution (2) and most of the measured width values were narrower than 10 m. With uniform constant widths of 1 m and 10 m (as two end members), cumulative frequency plots for MLE are compared to the reference simulation (Figure 12). As implied by Equation (3), the effects of the change in fracture zone width are clear in Figure 12. As fracture zones become wider, MLE at a specific depth decreases. Note that when the width varies within an order for a given probability density in (2), its impact is limited in a Shield environment.

### 5.2. Fracture Network Geometry

The fracture zone network model developed by *Srivastva* [2002] utilizes statistical tools for the downward propagation of surface lineaments according to geological and geomechanical principles. Due to its probabilistic nature, an infinite number of equally-probable fracture networks can be generated for a given set of surface lineaments. In addition, because a certain number of random surface features are generated to accommodate a given fracture density, this results in additional uncertainty for the generated subsurface fracture zones.

In this section, we use 100 equally-probable fracture zone network realizations to analyze the statistics and uncertainty in MLE distributions associated with the uncertainty in fracture zone location and geometry. Figure 13 shows the depth distributions of the probability for a certain location to belong to a fracture zone (hereafter termed as fracture zone probability, $\Pr_f$). The fracture zone probability generally decreases with depth and the randomly-generated surface lineaments are less probable to propagate to depth although some major fracture zones could penetrate downwards 1 km from surface.

For each network realization, finding a fracture zone in a given location follows a binomial distribution, with the fracture zone probability denoted as $\Pr_f$ and the matrix probability $1-\Pr_f$. The mean and standard deviation of the binomial distribution is given as $\Pr_f$ and $\Pr_f(1-\Pr_f)$, respectively, and thus the uncertainty is maximized when $\Pr_f = 0.5$ and it becomes certain when $\Pr_f = 0$ or 1. Uncertainty associated with major fracture zones honored by observed surface lineaments increases with depth due to the statistical variation of the down-dip propagation and thus red and yellow colors at shallower depths become green at deeper locations near major fracture zones as shown in Figure 13. For intermediate and minor fracture zones, their locations are highly uncertain when they are generated at surface but the uncertainty propagates only to shallow depths as they do not propagate to great depths.

MLE is simulated for 100 realizations of the fracture-zone networks. For each realization, hydraulic properties in the fracture zones and matrix for the reference simulations are maintained. At each node point, 100 MLE values are obtained from 100 realizations. Figure 14 shows distributions of 5th percentile (5th smallest out of 100 MLE values) and quartile variation coefficient (QVC) for 100 MLE values at each node, which is defined as:

$$\mathrm{QVC} = \frac{Q_3 - Q_1}{Q_3 + Q_1} \qquad (4)$$

where $Q_1$ and $Q_3$ represent first and third quartiles, respectively. For safety analyses, worst-case estimates need to be considered to ensure that the risk is minimal. The 5th percentile is a non-parametric worst-case estimate as 95% of events are expected to be safer than this level. The QVC is a non-parametric statistical measure for the relative degree of scattering of samples around a representative value such as the median. Order statistics such as percentiles and QVC can be more appropriate for a worst-case scenario and to quantify the uncertainty for a small number of samples when the population statistics are unknown.

The 5th percentile of computed MLE values at each node has similar distribution patterns as the MLE distribution for a single realization shown in Figure 6 (Figure 14a), indicating that the depth and the distance from major fracture zones are the most important factors affecting the safety of subsurface repositories. The results also indicate that the uncertainty associated with fracture-zone network geometry increases at shallower depths and near major fracture zones (Figure 14b).

## 6. Summary and Conclusions

The Canadian Shield is one of the most stable crystalline environments on Earth for hosting subsurface repositories. A hypothetical but realistic plutonic crystalline Shield setting was considered to analyze groundwater flow and lifetime expectancy, including topography, fracture zone network geometry, and hydraulic properties distributions. Lifetime expectancy in terms of mean and probability density was demonstrated to be a useful criterion for optimal repository locations as travel times from the repository to the biosphere become greater with more dilution, as lifetime expectancy increases.

Simulation results indicated that the lifetime expectancy generally increased with depth but it was also strongly influenced by the presence of permeable fracture zones: among others, fracture zone locations and the relationship between fracture-zone permeability and depth were most important for the given Shield setting. Although the first temporal moment of lifetime expectancy (MLE) could differentiate more stable regions from active groundwater flow regions, analysis of probability density for lifetime expectancy would be more desirable in order to capture the early arrival of contaminants in multiple pathway groundwater systems such as those containing a network of discrete fractures or fracture zones. It is worth noting that the analysis framework suggested by *Cornaton et al.* [2007] and this study are appropriate for preliminary siting of repositories, and thus, further careful multi-facetted analyses are required for the final evaluation of the protective performance of host rocks. Also, it cannot be over-emphasized that long-term evolution of groundwater flow systems as affected by Shield brines at depth, earthquakes, and glaciations could be important factors in assessing the protective performance of geologic barriers for the safe disposal of toxic wastes such as spent nuclear fuel.

## Acknowledgments

This research was supported by Ontario Power Generation (OPG), the Natural Sciences and Engineering Research Council (NSERC) of Canada and by the Swiss Research National Fund (Grant no. 2100-064927) to F. J. Cornaton. Additional funding was provided by grant 3-2-3 from Sustainable Water Resources Research Center of the 21$^{st}$ Century Frontier Research Program of Korea.

**Figure Captions**

**Figure 1.** Aerial photo of the sub-regional domain.

**Figure 2.** (a) A single realization of fracture-zone networks in the sub-regional domain and (b) fracture-zone geometry mapped onto model elemental faces, along with fracture-zone permeability [m$^2$] as a function of depth.

**Figure 3.** (a) Fracture-zone permeability data with depth measured at the WRA and (b) a probabilistic model developed to describe the permeability-depth relationship [*Normani et al.*, 2007].

**Figure 4.** Simulated hydraulic head distribution in the domain.

**Figure 5.** Darcy flux distribution at four different elevations of the domain of 100, -100, -300, and -500 m above sea levels (about 300, 500, 700, and 900 m below ground surface).

**Figure 6.** Mean lifetime expectancy distribution at four different elevations of the domain. P1-P4 represents 100m by 200m by 1km of volume for computation of probability density in lifetime expectancy.

**Figure 7.** Cumulative relative frequency plots for mean lifetime expectancy at four elevations of the domain in Figure 6.

**Figure 8.** Probability density of lifetime expectancy for P1 – P4 in Figure 6.

**Figure 9.** Mean lifetime expectancy distributions with (a) a uniform fracture-zone permeability model and (b) a probabilistic decrease with depth for fracture-zone permeability.

**Figure 10.** Comparison of cumulative frequency plots for MLE with (a) a uniform fracture-zone permeability model and (b) a probabilistic decrease with depth for fracture-zone permeability, to the reference result with the average deterministic model.

**Figure 11.** Comparison of cumulative frequency plots for MLE with uniform fracture-zone porosity ($n_F$) of (a) 0.3 % and (b) 3 %, to the reference result with Equation (1).

**Figure 12.** Comparison of cumulative frequency plots for MLE with uniform fracture-zone width of (a) 1 m and (b) 10 m, to the reference result with Equation (2).

**Figure 13.** Fracture-zone probability distribution in the domain.

**Figure 14.** Distribution of (a) 5th percentile and (b) quartile variation coefficient for 100 MLE values from 100 fracture-zone network realizations.

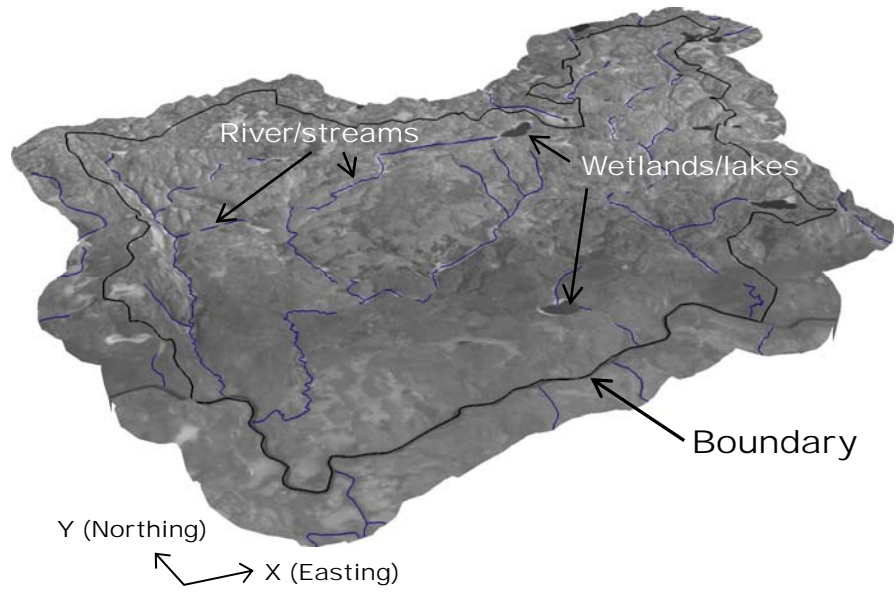

**Figure 1.** Aerial photo of the sub-regional domain.

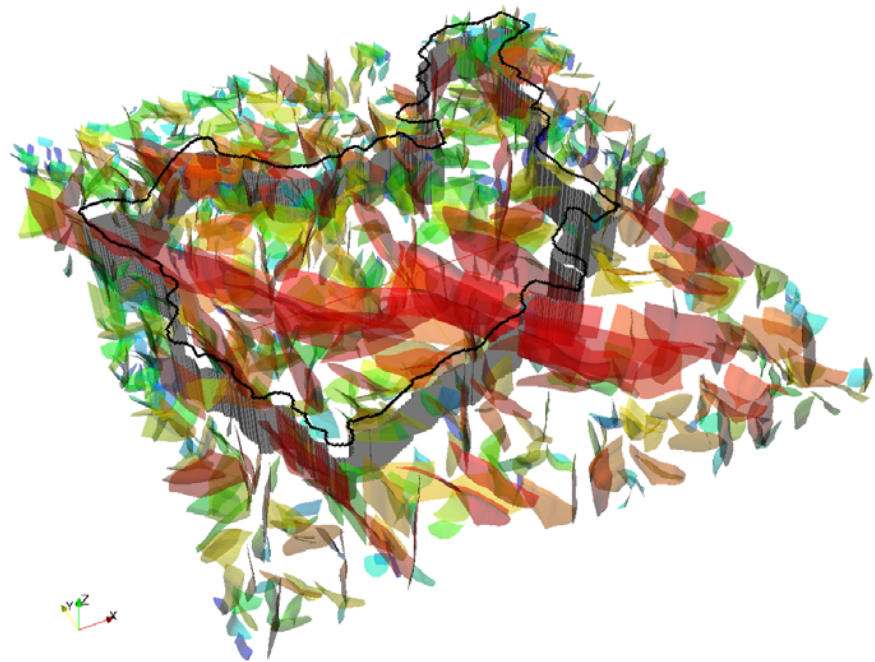

(a)

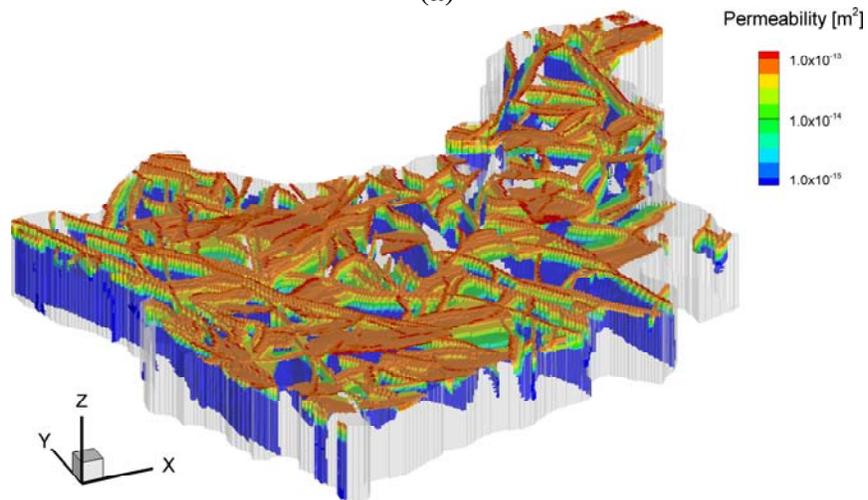

(b)

**Figure 2.** (a) A single realization of fracture-zone networks in the sub-regional domain and (b) fracture-zone geometry mapped onto model elemental faces, along with fracture-zone permeability [m$^2$] as a function of depth.

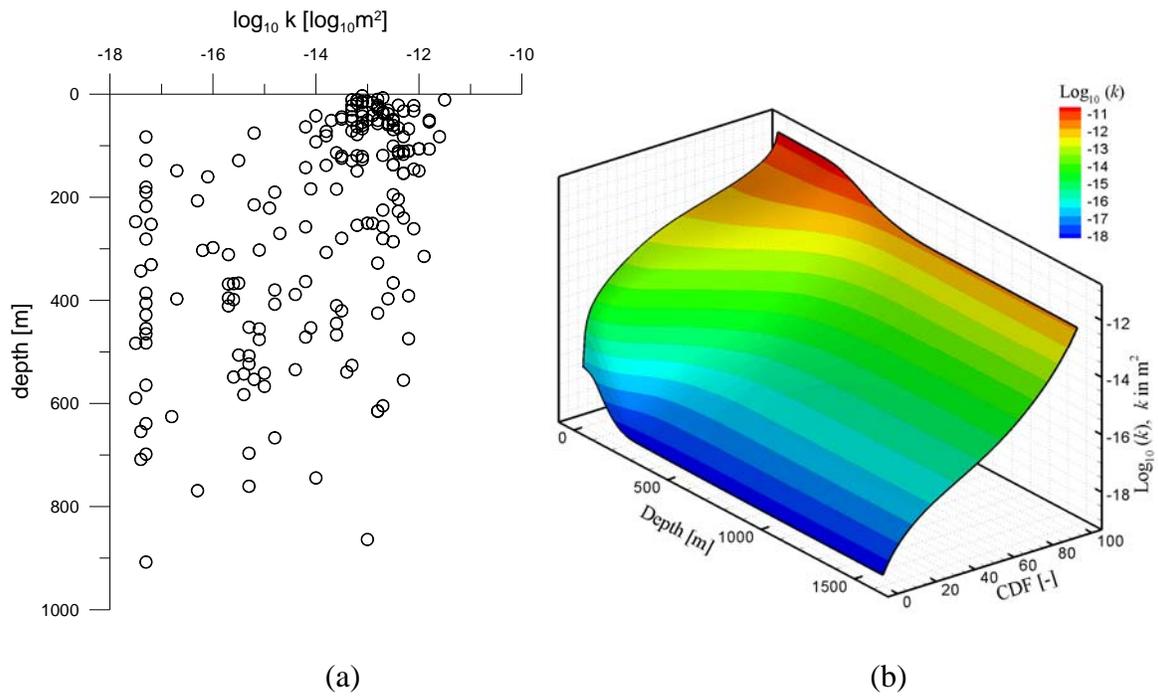

**Figure 3.** (a) Fracture-zone permeability data with depth measured at the WRA and (b) a probabilistic model developed to describe the permeability-depth relationship [*Normani et al.*, 2007].

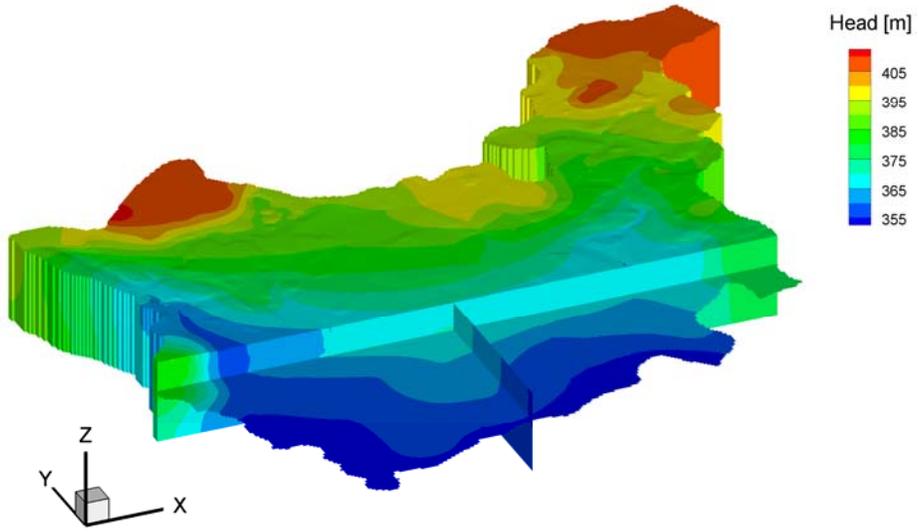

**Figure 4.** Simulated hydraulic head distribution in the domain.

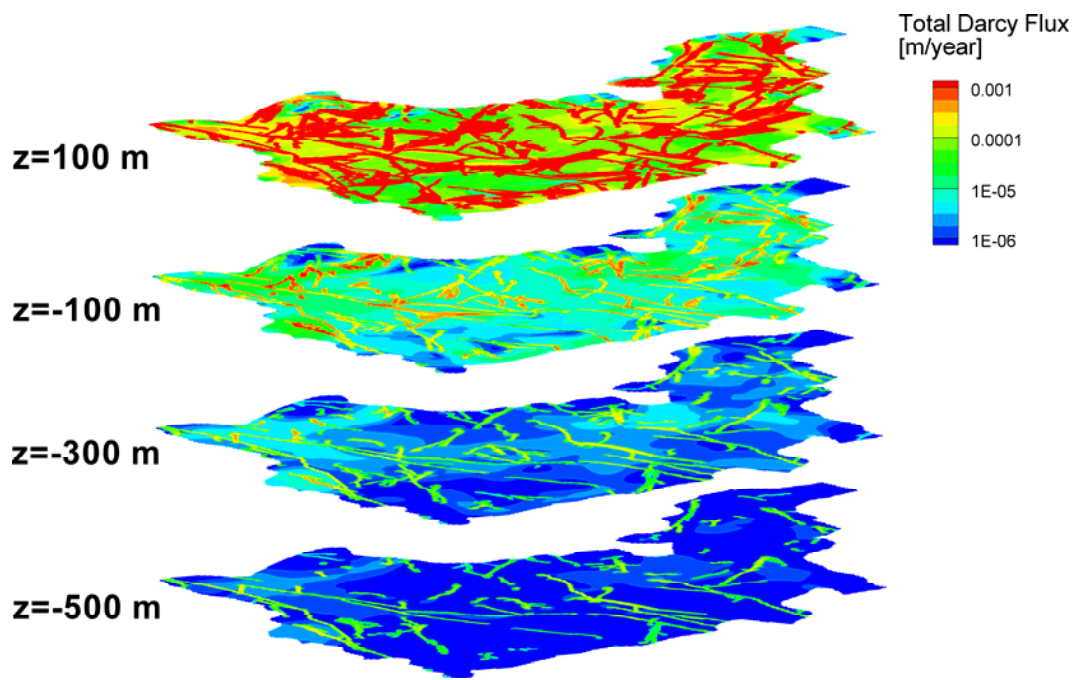

**Figure 5.** Darcy flux distribution at four different elevations of the domain of 100, -100, -300, and -500 m above sea levels (about 300, 500, 700, and 900 m below ground surface).

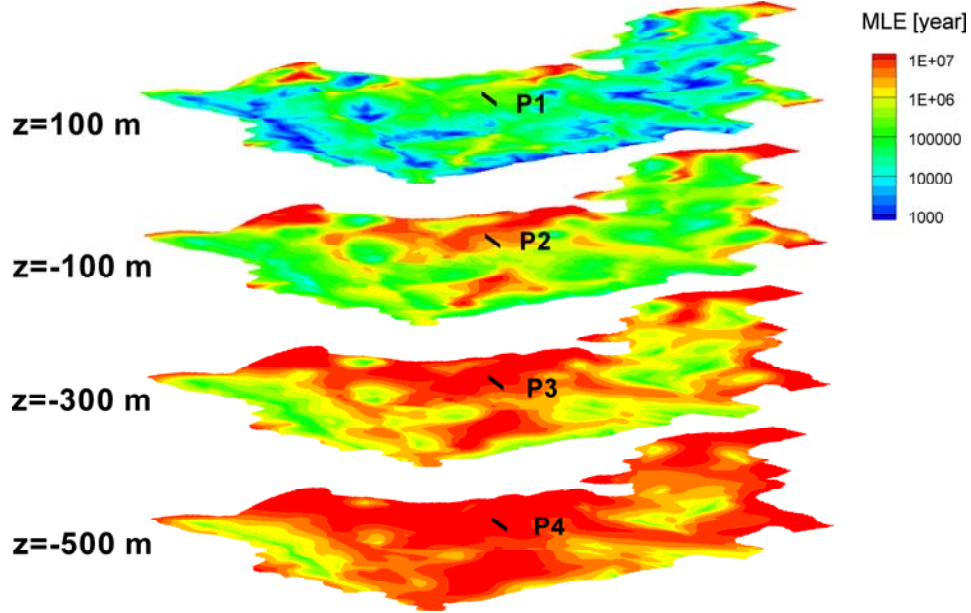

**Figure 6.** Mean lifetime expectancy distribution at four different elevations of the domain. P1-P4 represents 100m by 200m by 1km of volume for computation of probability density in lifetime expectancy.

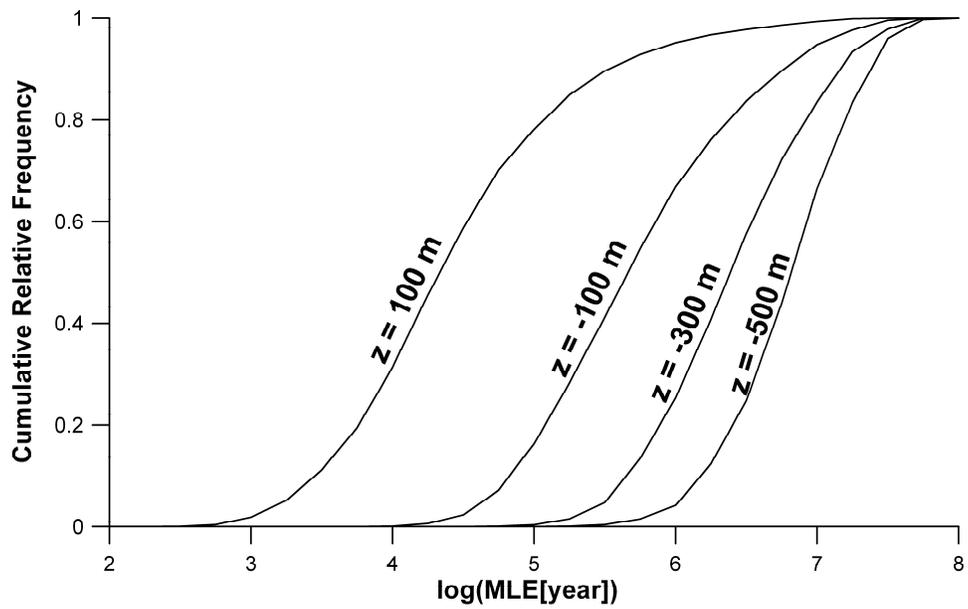

**Figure 7.** Cumulative relative frequency plots for mean lifetime expectancy at four elevations of the domain in Figure 6.

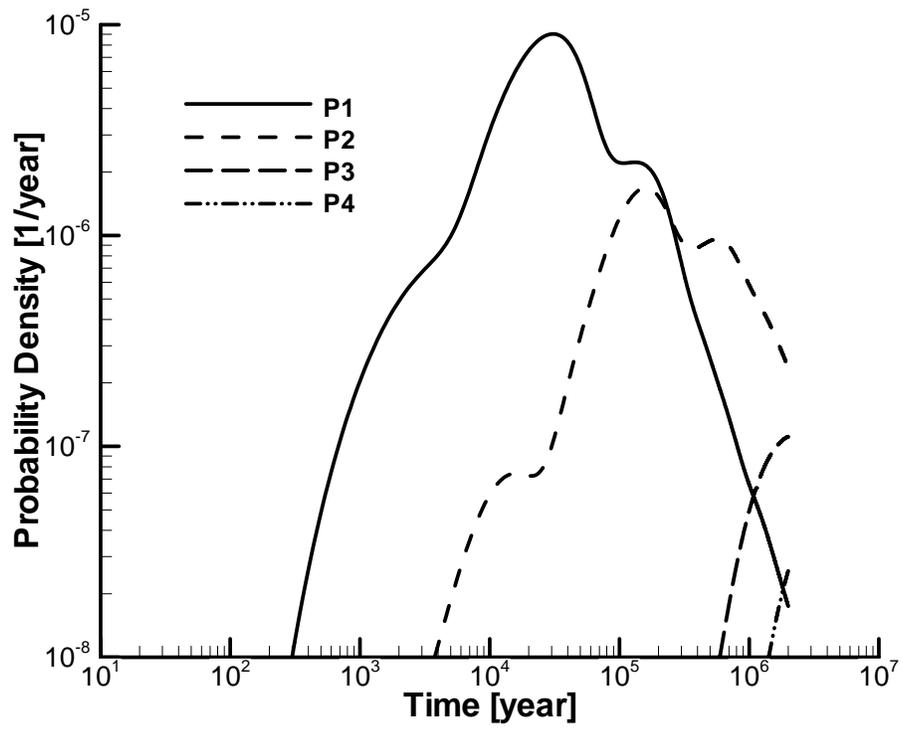

**Figure 8.** Probability density of lifetime expectancy for P1 – P4 in Figure 6.

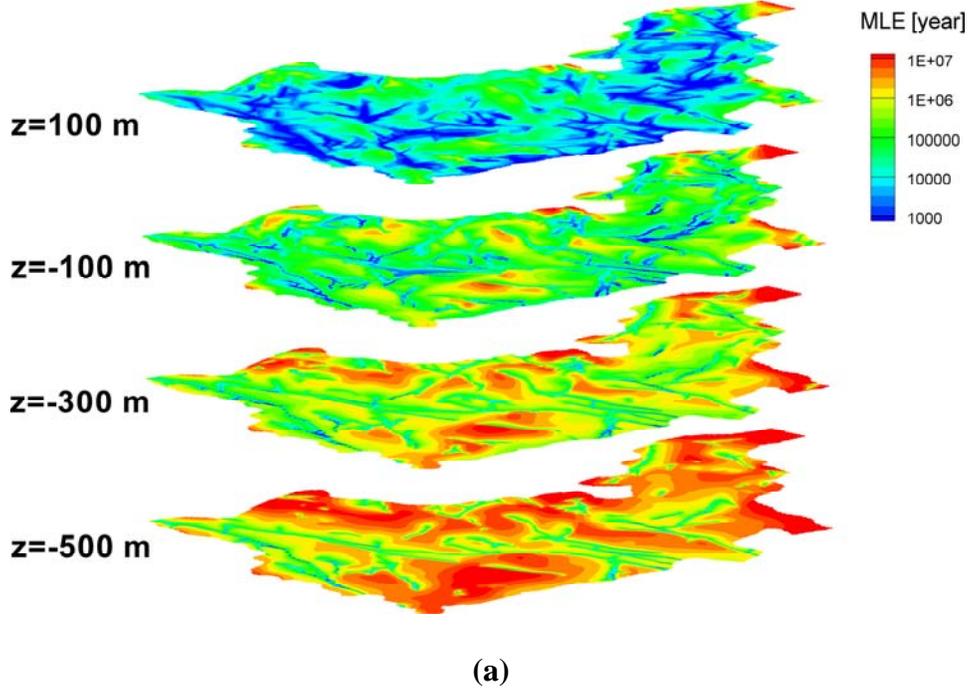

(a)

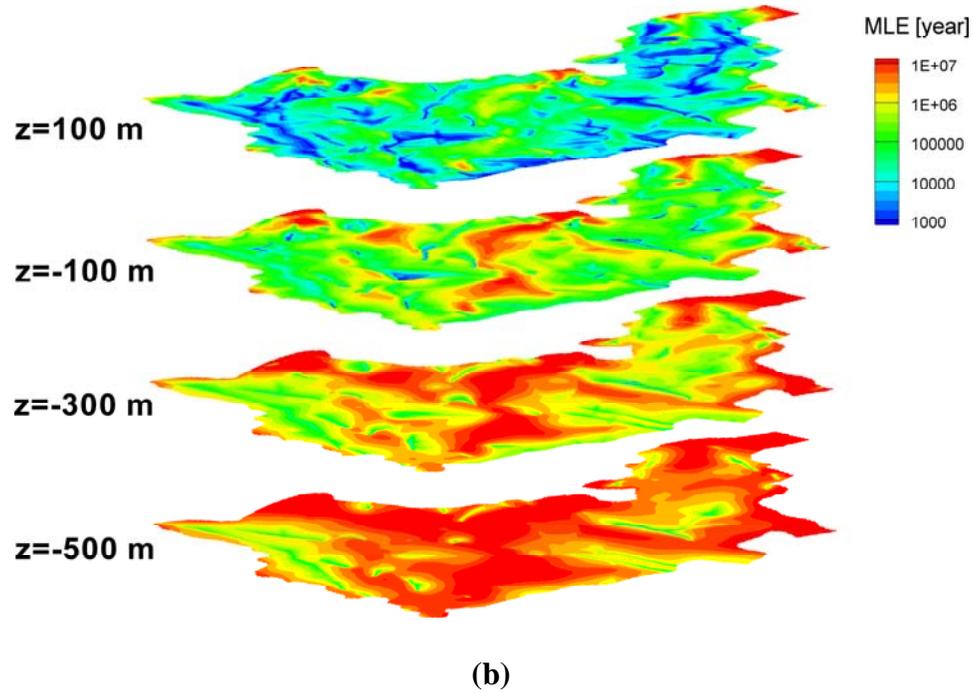

(b)

**Figure 9.** Mean lifetime expectancy distributions with (a) a uniform fracture-zone permeability model and (b) a probabilistic decrease with depth for fracture-zone permeability.

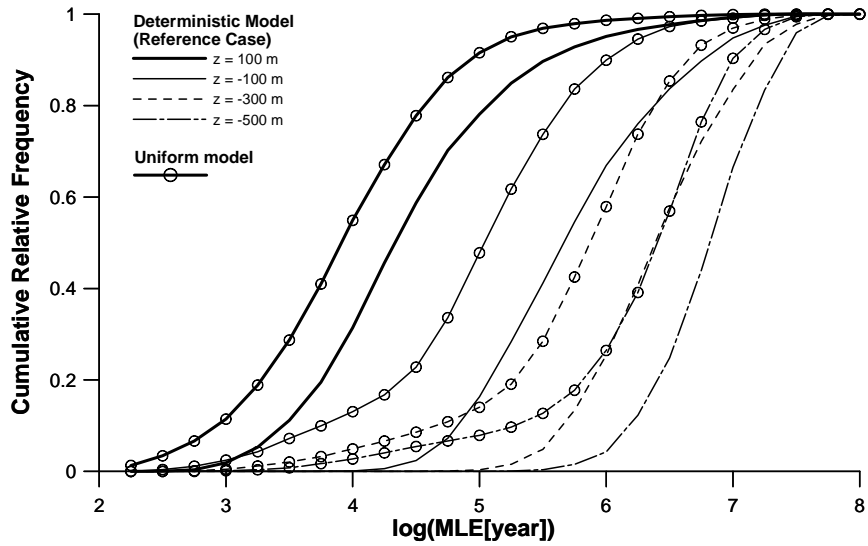

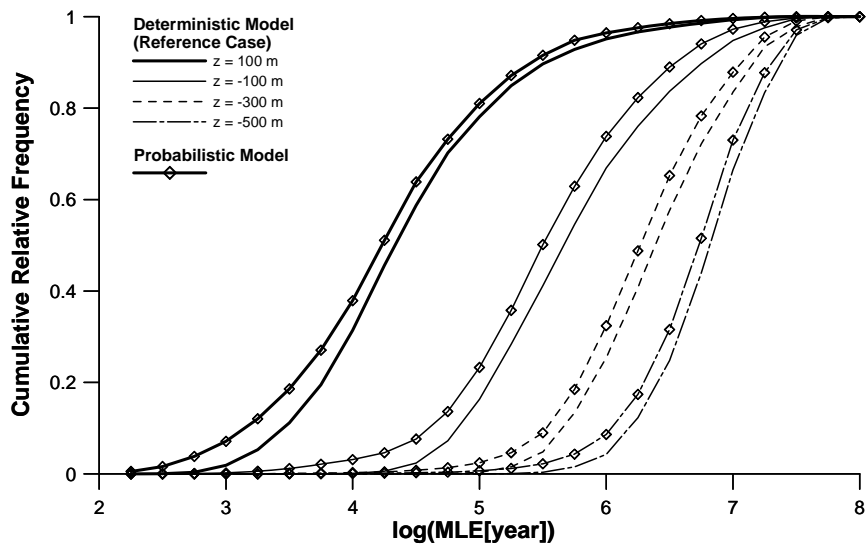

**Figure 10.** Comparison of cumulative frequency plots for MLE with (a) a uniform fracture-zone permeability model and (b) a probabilistic decrease with depth for fracture-zone permeability, to the reference result with the average deterministic model.

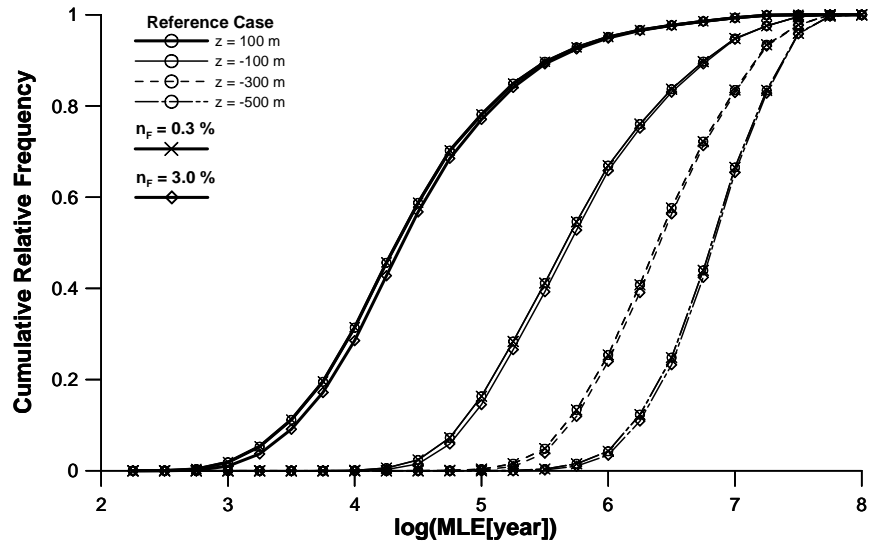

**Figure 11.** Comparison of cumulative frequency plots for MLE with uniform fracture-zone porosity ($n_F$) of (a) 0.3 % and (b) 3 %, to the reference result with Equation (1).

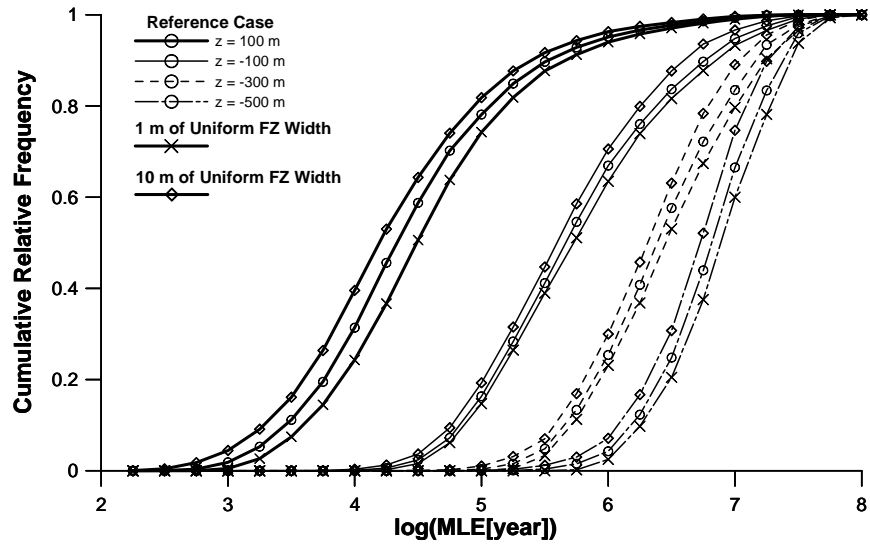

**Figure 12.** Comparison of cumulative frequency plots for MLE with uniform fracture-zone width of (a) 1 m and (b) 10 m, to the reference result with Equation (2).

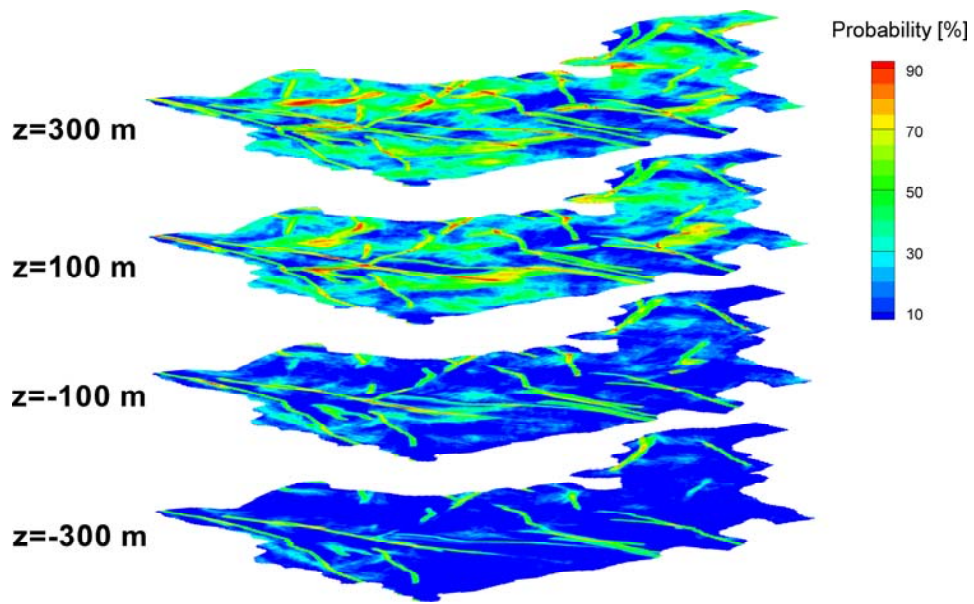

**Figure 13.** Fracture-zone probability distribution in the domain.

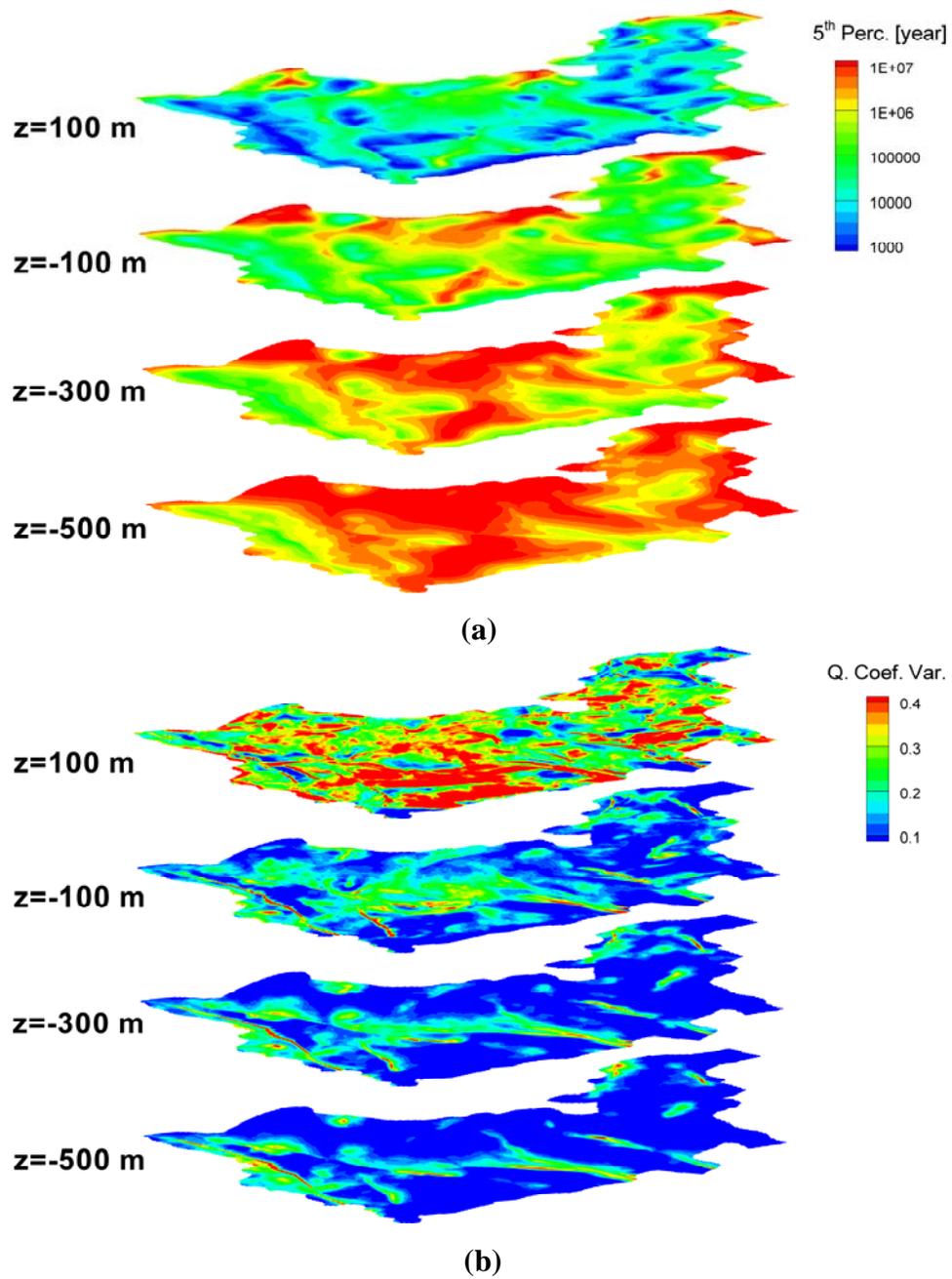

**Figure 14.** Distribution of (a) 5th percentile and (b) quartile variation coefficient for 100 MLE values from 100 fracture-zone network realizations.